\documentclass[final,5p,times,twocolumn]{elsarticle}

\usepackage{graphicx}%
\usepackage{multirow}%
\usepackage{amsmath,amssymb,amsfonts}%
\usepackage{amsthm}%
\usepackage{mathrsfs}%
\usepackage[title]{appendix}%
\usepackage{xcolor}%
\usepackage{textcomp}%
\usepackage{manyfoot}%
\usepackage{tabularx,booktabs}%
\usepackage{algorithm}%
\usepackage{algorithmicx}%
\usepackage{algpseudocode}%
\usepackage{listings}%

\usepackage{subcaption}
\usepackage[export]{adjustbox}
\journal{Image and Vision Computing}

\begin{document}

\begin{frontmatter}


 \title{A New Multi-Picture Architecture for Learned Video Deinterlacing and Demosaicing \\ with Parallel Deformable Convolution and Self-Attention Blocks\tnoteref{titlelabel}}
 \tnotetext[titlelabel]{This paper extends early results reported in our ICIP 2022 paper~\cite{ji2022multi}.}
 \author[1]{Ronglei Ji}
 \ead{rji19@ku.edu.tr}
 \author[1]{A. Murat Tekalp\corref{cor1}}
 \ead{mtekalp@ku.edu.tr}

 \cortext[cor1]{Corresponding author.}
 \affiliation[1]{organization={Department of Electrical and Electronics Engineering, Koç University},
             city={Istanbul},
             postcode={34450},
             country={Turkey}}




\begin{abstract}
Despite the fact real-world video deinterlacing and demosaicing are well-suited to supervised learning from synthetically degraded data because the degradation models are known and fixed, learned video deinterlacing and demosaicing have received much less attention compared to denoising and super-resolution tasks. We propose a new multi-picture architecture 
for video deinterlacing or demosaicing by aligning multiple supporting pictures with missing data to a reference picture to be reconstructed, benefiting from both local and global spatio-temporal correlations in the feature space using modified deformable convolution blocks and a novel residual efficient top-$k$ self-attention (kSA) block, respectively. Separate reconstruction blocks are used to estimate different types of missing data. 
Our extensive experimental results, on synthetic or real-world datasets, 
demonstrate that the proposed novel architecture provides superior results that significantly exceed the state-of-the-art for both tasks in terms of PSNR, SSIM, and perceptual quality. Ablation studies are provided to justify and show the benefit of each novel modification made to the deformable convolution and residual efficient kSA blocks. Code is available: https://github.com/KUIS-AI-Tekalp-Research-Group/Video-Deinterlacing.

\end{abstract}




\end{frontmatter}


\section{Introduction}
Mosaicing refers to a color sub-sampling scheme using a color filter array (CFA) pattern that is universally employed in all single-sensor digital still and video cameras. 

Interlacing is a spatio-temporal sampling scheme, consisting of odd and even fields, that was introduced to strike a balance between spatial and temporal resolution because analog TV broadcasting did not support a bandwidth to deliver progressive video at a high enough frame rate. The interlaced scan doubles the temporal rate at the same bandwidth at the expense of vertical spatial resolution resulting in better visual perception of scenes with fast motion. Even though almost all digital cameras today are progressive, deinterlacing remains as a problem of interest to display interlaced broadcast on progressive displays as well as conversion of existing interlaced and telecined catalogue content for progressive broadcast.

Demosaicing and deinterlacing are commonly used in the~image processing pipeline for consumer video. Yet, the~industry employs simple methods for demosaicing, deinterlacing, telecine, and 50-to-60 Hz and vice versa conversion.
Over the~years, many conventional solutions have been proposed by the academia for demosaicing~\cite{Losson2010}, intra-field deinterlacing~\cite{kim2007novel,yoo2002direction}, inter-field deinterlacing~\cite{kwon2003deinterlacing}, and spatio-temporal super-resolution~\cite{shechtman2005space}. However,~these solutions do not employ 
learned image/video priors; hence, introduce blurring or disturbing color artifacts, edge artifacts, and/or flicker.

Real-world demosaicing and deinterlacing are well-suited for supervised learning from synthetic data, because the forward (degradation) model is pre-determined and is fixed for these upsampling problems. The~forward model for both problems can be expressed as:
\begin{flalign}
\textbf{y}=\textbf{Sx} + \textbf{v}&&
   \label{eq:down}
\end{flalign}
where {\bf x} is the ideal progressive video, {\bf y} denotes the subsampled video, {\bf v} is observation noise, and {\bf S} is the subsampling matrix defined by the interlace or known CFA pattern. 

The~literature on learned video demosaicing and deinterlacing benefiting from {\it both local and global spatio-temporal correlations} is scarce. Moreover, the literature treats demosaicing and deinterlacing as separate problems although they share a common forward model~(\ref{eq:down}).
In this paper, we propose a single novel multi-picture architecture (for both problems) that integrates both local and global spatio-temporal features extracted by modified deformable convolution blocks and efficient residual top-$k$ self-attention blocks, respectively, and employs separate reconstruction blocks to estimate the missing pixels. The~rest of this paper is organized as: We~review related works and our main contributions in Section~\ref{related}. Section~\ref{method} introduces the proposed common multi-picture architecture. Section~\ref{expr} presents extensive experiments to show that our model yields state-of-the-art results and generalizes better than all existing methods for both problems when trained and tested on different datasets. Section~\ref{conc} concludes the paper.

\section{Related works and contributions}
\label{related}
We review related works on learned demosaicing and deinterlacing  in Section~\ref{related_1} and Section~\ref{related_2}, respectively. Prior works on deformable convolutions and integration of local and global features are reviewed in Section~\ref{related_3} and Section~\ref{related_SA}, respectively, in order to more clearly identify our novel contributions on these fundamental blocks. Section~\ref{related_5} summarizes the~novelty and contributions of this paper.

\subsection{Learned image and video demosaicing}
\label{related_1}
Deep learned models provide important quantitative and visual performance gains over conventional solutions. 
Early approaches for still images employed multi-model~\cite{tan2018demosa} or multi-stage \cite{wang2020ntsdcn} deep convolutional networks (CNN) to reconstruct and refine the G, R and B channels. Zhang {\it et al.}~\cite{zhang2021plug} employed a deep CNN denoiser prior for plug-and-play restoration. Feng {\it et al.}~\cite{feng2021mosaic} applied convolution-attention network to multispectral demosaicing. Xu {\it et al.}~\cite{xu2020joint} jointly combined demosaicing with superresolution problem. Sharif {\it et al.}~\cite{a2021beyond} introduced a learning based method to jointly tackle demosaicing and denoising problems.
Zhang {\it et al.}~\cite{zhang2022deep} proposed a spatially adaptive convolutional network by feeding the color filter array (CFA) pattern to the~network, which inspired us to propose the separate reconstruction blocks in this paper.

Multi-picture demosaicing that take spatio-temporal color correlations into account have only very recently been considered for burst image and video demosaicing. Kokkinos and Lefkimmiatis~\cite{kokkinos2019iterative} applied convolutional iterative network on burst images by aligning each supporting picture to the reference using enhanced correlation coefficient. Ehret {\it et al.}~\cite{ehret2019joint} employed an inverse compositional algorithm to register frames when performing fine-tuning on burst images. Dewil {\it et al.}~\cite{dewil2023video} proposed recurrent networks with motion compensation for joint video denoising and demosaicing.

In this paper, we propose a common multi-picture architecture for video demosaicing and deinterlacing that
extracts both local and global spatio-temporal features by  using modified deformable convolutions and efficient top-$k$ self attention, respectively, in parallel. These features are additively integrated. We~also show that using separate reconstruction blocks for R, G, and B channels provides better results than conditioning on the CFA pattern as done in~\cite{zhang2022deep}.

\subsection{Learned deinterlacing}
\label{related_2}
Early works either consider only intra-frame deinterlacing~\cite{zhu2017real,akyuz2020deep} or do not employ feature-level alignment of supporting fields~\cite{bernasconi2020deep, zhaointerlace}; hence, their results are suboptimal in the presence of large motion. There are few recent works that consider feature-level field alignment using deformable convolutions for the deinterlacing task. One of the first works was by us~\cite{ji2021learned}. Other papers that use deformable convolutions for feature-level field alignment for deinterlacing have concurrently appeared~\cite{liu2021spatial,zhao2022multi}. A recent paper~\cite{gao2023revitalizing} utilized optical flow guided deformable convolution for feature alignment but did not use self-attention (SA) features.  Another recent paper~\cite{song2023transformer} proposed a transformer based deinterlacing network to take advantage of inter-frame dependencies by incorporating SA only.

In our conference paper~\cite{ji2022multi}, we proposed combining features aligned by deformable convolution blocks with those computed by global SA layers, and then processing additively fused features by separate reconstruction modules for the first time.
This~paper extends and improves the preliminary results in~\cite{ji2022multi}. In particular, we replace the SA  blocks with an efficient residual top-$k$ SA block for global feature processing. We also modify the deformable convolution blocks in~\cite{ji2022multi} by removing the redundant residual connection and adding a convolution layer between the two deformable convolution layers to obtain better performance. We elaborate on the novelty of our paper in more detail in Section~\ref{related_5}.

\subsection{Deformable convolution for video processing}
\label{related_3}
\par Deformable convolution, with a larger and more flexible receptive field compared to regular convolution, was first proposed~for object detection and semantic segmentation~\cite{dai2017deformable}. In~deformable convolution, pixel positions under the~kernel are displaced according to learned offsets $\Delta p_{(i,j)}$. Hence, the~deformable convolution can be expressed as:
\begin{flalign}
y(p_{(m,n)})=\sum_{(i,j)} w(i,j) \cdot x(p_{(m-i,n-j)}+\Delta p_{(i,j)})&&
\label{2}
\end{flalign}
where $p_{(m,n)}$ is the current pixel position, $x(p_{(m,n)})$ is the pixel at position $p_{(m,n)}$, $w(i,j)$ are kernel weights, and for a $3\times3$ deformable convolution $i,j \in (-1,0,1)$.

It~has~later been adopted for video processing tasks, such as video super-resolution (VSR)~\cite{tian2020tdan, wang2019edvr, ying2020deformable}, frame rate up-conversion~\cite{cheng2021multiple}, and next-frame prediction~\cite{yilmaz2021dfpn}. Although different aligning methods exist for other tasks~\cite{zhang2022distinguishing,kumari2021feature}, temporal alignment of supporting pictures with a reference picture at the~feature level using deformable convolution without using optical flow was first proposed in TDAN~\cite{tian2020tdan}. This was extended to the pyramid, cascading, and deformable convolution alignment sub-network in EDVR~\cite{wang2019edvr} for the VSR task. 

In this paper, we modified the offset estimation method in deformable convolutions tailoring it to demosaicing and deinterlacing tasks, and proposed a modified deformable convolution~(DfConv) block for feature-level picture alignment.

\subsection{Integrating local and global features via self-attention}
\label{related_SA}
Given the success of non-local image processing and the~ability of self-attention (SA) to exploit long range interactions~\cite{chen2022improving}, there is growing interest in using SA layers either as stand-alone primitives or to augment ConvNets for image processing. Attention~\cite{li2023external,zhu2022fine} augmented convolutional networks combining both convolutional and SA layers by concatenating respective feature maps has been proposed~\cite{bello2019attention}. 
However, the~memory requirement and computational complexity of modeling global dependency by dot-product attention are quadratic with respect to picture size, which prohibits its application to high-resolution images and large videos. 

Recently, several works have emerged proposing efficient SA implementations~\cite{tay2020efficient}.
Shen {\it et al.} proposed an efficient implementation of SA with linear complexity~\cite{shen2021efficient}.
A later work~\cite{xia2022efficient} further incorporates a Gaussian approximation to decompose the exponential function of queries and keys in order to change the order of multiplications.

Both~\cite{shen2021efficient} and~\cite{xia2022efficient} utilize all tokens in matrix multiplications, which contributes to the computation cost as well as introducing a noise effect due to irrelevant tokens. Recently, Wang {\it et al.}~\cite{wang2021kvt} proposed an efficient $k$-NN self-attention to keep only the top $k$ similar tokens instead of all tokens in the attention weight map in order to benefit from long range dependencies while filtering out irrelevant tokens.

In this paper, we combine these ideas with some improvements to propose the residual efficient kSA blocks: i) we show that the kSA block with a residual connection yields better performance than without a residual connection, ii)~different from~\cite{bello2019attention}, we show that combining features computed by deformable convolution and kSA blocks 
by addition rather than concatenation yields better performance, iii) we propose an efficient top-$k$ SA by changing multiplication order to reduce training and test time while preserving the performance.

\subsection{Novelty and advancing the state of the art}
\label{related_5}
This paper advances the state-of-the-art in multi-picture learned video demosaicing and deinterlacing as follows:
\par 1) We propose a common multi-picture architecture for video demosaicing and deinterlacing (since both are video upsampling problems) featuring modified deformable convolution layers and efficient residual top-$k$ SA layers to additively combine local and global features, respectively, and separate reconstruction blocks for each channel to be upsampled.
\par 2) We modify the DfRes blocks proposed in~\cite{ji2022multi} by removing the redundant residual connection and inserting a regular convolution layer in between the two deformable convolution layers (Df). We show that the modified Deformable Convolution blocks (DfConv) yield better results than  DfRes blocks.
\par 3) We replace the SA block in~\cite{ji2022multi}
with novel efficient residual top-$k$ SA block. We show that changing multiplication order can reduce the training and test time to about half compared to~\cite{ji2022multi}
while retaining performance, and that adding a residual connection in the top-$k$ SA block yields a remarkable boost in the performance.
\par 4) We demonstrate that adding the features aligned by the~deformable convolution layers and self-attention layers (in parallel) instead of concatenating them as proposed in~\cite{bello2019attention} yields better results.
\par 5) We show that using separate reconstruction blocks for each picture to be reconstructed according to its particular subsampling pattern yields superior performance compared to using a single reconstruction block for all picture types.
\par 6) Our results are the state-of-the-art for both demosaicing and deinterlacing tasks. We provide ablation studies on the value of $k$ 
and also to show how much improvement comes from each of the~proposed modifications, as well as extensive results to show that our architecture provides superior generalization performance over competing methods for both tasks when trained and tested on different datasets.


\begin{figure*}[t!]
  \centering
  \subfloat[]{
  \includegraphics[scale=0.65]{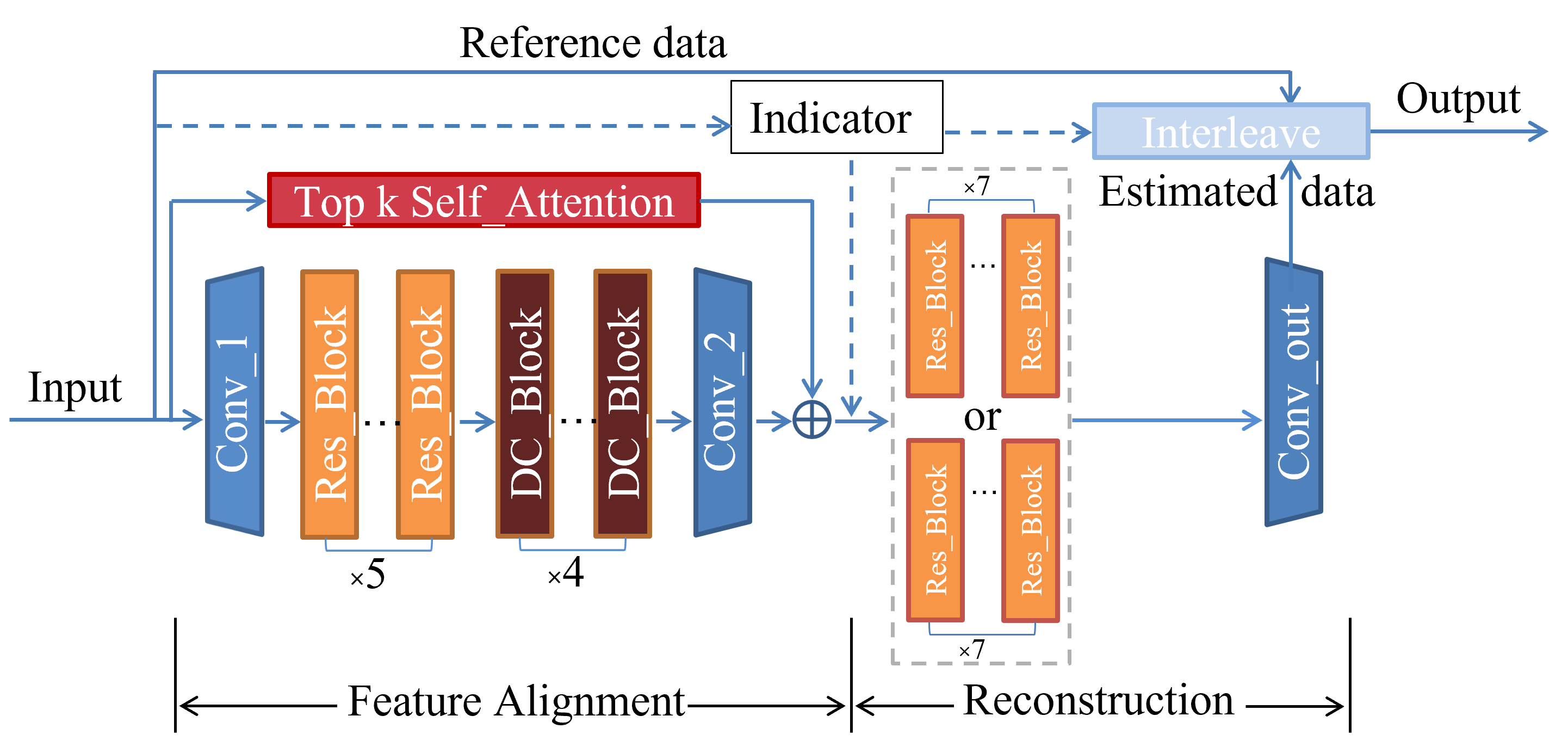}
  }
    \\
\centerline{\subfloat[]{
    \includegraphics[scale=0.65]{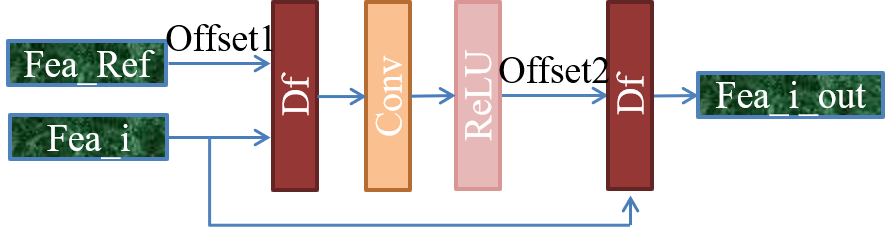} 
    }
    \hspace{30pt}
  \subfloat[]{
    \includegraphics[scale=0.65]{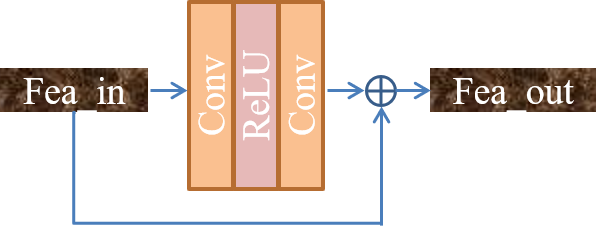}
    }
    }
\caption{The proposed multi-picture network architecture. (a) The overall architecture, where DfConv blocks and top-$k$ self-attention block extract and align local and global spatio-temporal features, respectively. The aligned features are integrated additively. The reconstruction stage consists of separate reconstruction blocks (two 7 $\times$ Res\_Blocks for deinterlacing and three 7 $\times$ Res\_Blocks for demosaicing) according to the subsampling pattern. The solid arrows show data flow and the dashed arrows depict indicator signaling.
(b) Block diagram of a DfConv block (DC\_Block, brown boxes in (a)), where Fea\_i denotes input features and Fea\_i\_out denotes the corresponding output features, which will be input to the next DfConv block. (c) Block diagram of a standard residual block (Res\_Block, orange boxes~in~(a)).}
\label{fig} 
\end{figure*}

\section{Proposed architecture}
\label{method}
We first present the rationale and an overview of our proposed architecture in Section~\ref{netarch}. We introduce the main blocks of the proposed architecture: deformable convolution blocks, efficient residual top-k SA block, and separate reconstruction blocks, in Section~\ref{defconv} to Section~\ref{seprecon}, respectively.

\subsection{Rationale and overview}
\label{netarch}
The rationale for a common architecture for both tasks is that they are both video upsampling problems (with different but fixed subsampling patterns), which would benefit from spatio-temporal modeling of features. Hence, a common architecture modeling both local and global spatio-temporal interactions using deformable convolutions and self attention, respectively, and separate reconstruction modules for different fields/channels should serve both tasks well.

The proposed multi-field video upsampling network, depicted in Fig.~\ref{fig}(a), consists of two stages: i) feature alignment and integration stage, where local features of supporting pictures are computed and aligned with those of the current reference picture using modified deformable convolution blocks, and then integrated additively with the global features computed by efficient residual top-$k$ self-attention layer in parallel, and ii)~reconstruction stage with separate reconstruction blocks (two for deinterlacing and three for demosaicing) to estimate missing pixels in each subsampled component.

The input to the alignment stage consists of an odd number of pictures centered about the reference picture. 
In the deinterlacing task, for example, if we were to estimate an even field $N_E$ using 5~fields, then the reference field is $N_O$ and the~input fields would be $(N-2)_O, (N-1)_E, N_O, (N+1)_E, (N+2)_O$. Although the reconstructed fields $(N-1)_O$ and $(N-2)_E$ would be available at test time since they are in the past of the current field to be estimated, we chose not to use them in order to avoid potential error propagation.
In the demosaicing task, we have successive R, G, B pictures with missing pixels as input. We use an indicator flag to indicate the type (i.e., odd/even or R/G/B) of the picture with missing lines/pixels according to the~subsampling pattern.

It is well-known that deformable convolutions can achieve excellent temporal alignment of local features in video restoration and SR tasks~\cite{tian2020tdan,wang2019edvr}. It is also well-known that self attention (SA) expands the receptive field to the entire picture. However, the combination of two has been seldom used to integrate local and global features for image/video upsampling tasks. Our previous work~\cite{ji2022multi} uses the standard vanilla SA, which involves all tokens into the calculations, including a number of irrelevant tokens in high resolution image/video processing that inject noise into the attention results. It has been observed that this negatively affects the training process~\cite{wang2021kvt}. In this work, we employ top-$k$ SA (kSA) block, which selects the most relevant top-$k$ (not all) tokens from the attention feature map, to enhance the performance of feature registration and integration. We also noticed that adding a residual connection to top-$k$ SA block provides a significant performance improvement.
The local and global feature tensors processed by the DfConv and kSA block branches, respectively, in parallel are added together to form the~integrated feature tensor. This architecture is different from previous use of SA to augment ConvNets~\cite{bello2019attention} in computer vision tasks, where they connect convolutional and SA layers in series and concatenate respective features. We demonstrate the~effectiveness of our proposed strategy in Section~\ref{expr}.


The aligned and integrated feature tensor is input to the reconstruction stage, where it is processed by one of the~separate reconstruction blocks as determined by the indicator~flag. Each~reconstruction block has seven regular residual blocks. The~last Conv\_out layer maps the processed feature tensor to three~channels to generate the reconstructed pixels. Finally, the estimated pixels are interleaved with the available reference pixels for deinterlacing task or with the other two estimated channels for demosaicing task to reconstruct the output frame. 

\subsection{Modified deformable convolution blocks}
\label{defconv}
Each input picture sequence with missing data is first mapped to a 64-channel feature tensor via the convolution layer Conv\_1 and next processed by five regular residual blocks to refine the~features. We employ deformable convolution (DfConv) blocks (shown in Fig.~\ref{fig}(a) brown boxes labeled as DC\_Block), to align spatio-temporal features due to their flexible receptive fields. The processing chain is depicted in Fig.~\ref{fig}(a), where we use 5$\times$ standard Res\_Blocks for feature extraction and 4$\times$ DfConv blocks for feature alignment.
In the following, we propose a new DfConv block structure that yields better performance than using the standard deformable convolution blocks~\cite{tian2020tdan} and deformable convolution residual blocks (DfRes)~\cite{ji2022multi} for feature alignment. 

In the literature, offsets for deformable convolution are learned by applying regular convolutions directly to concatenated input feature maps \cite{dai2017deformable,tian2020tdan,cheng2021multiple,ying2020deformable}. Clearly, the~quality of learned offsets directly affects the performance of deformable convolution. We found that learning the offsets by using deformable (rather than standard) convolutions with a refining convolution layer (different from DfRes~\cite{ji2021learned} and DfRes+SA~\cite{ji2022multi} blocks) 
yields better performance.

Our proposed deformable convolution block (DfConv), shown in Fig.~\ref{fig}(b), employs two deformable convolution layers~(Df) with a regular convolution layer (Conv) between them, to align features of supporting pictures with those of the~reference picture. In the first Df layer, we treat features of the~reference picture (Fea\_Ref) as offsets (offset1) and perform Df on features of each supporting picture (Fea\_i) separately as described by the pseudocode:
\begin{flalign}
&offset1 = Fea\_Ref
\label{eq:Dflayer11} &\\
&interm\_out = Df(offset1, Fea\_i)
\label{eq:Dflayer12}&
\end{flalign}
The~results (interm\_out) are refined through a regular convolution layer (Conv) and ReLU activation, and then fed into the~second Df layer as second offsets (offset2) to produce the~aligned output features (Fea\_i\_out) for $\textit{i}^{th}$ supporting picture through one DfConv block as expressed by the pseudocode:
\begin{flalign}
&offset2 = ReLU(Conv(interm\_out))
\label{eq:Dflayer21} &\\
&Fea\_i\_out = Df(offset2, Fea\_i)
\label{eq:Dflayer22} &
\end{flalign}

There are a total of 4 DfConv blocks, where each time, we align one of the supporting pictures to the reference picture. Here, the difference from our prior work~\cite{ji2022multi} is that, firstly, we remove the redundant residual connection which we observe degrades the performance, and secondly, we add a regular convolution layer in between two Df layers, which refines the offsets. The effectiveness of these two modifications is demonstrated by an ablation study (shown in the top row of Table~\ref{ablation}).


\subsection{Efficient residual top-$k$ self-attention block}
\label{selfattention}
\par
\begin{figure}[t!]
\subfloat[]{
\includegraphics[scale=0.45,valign=t]{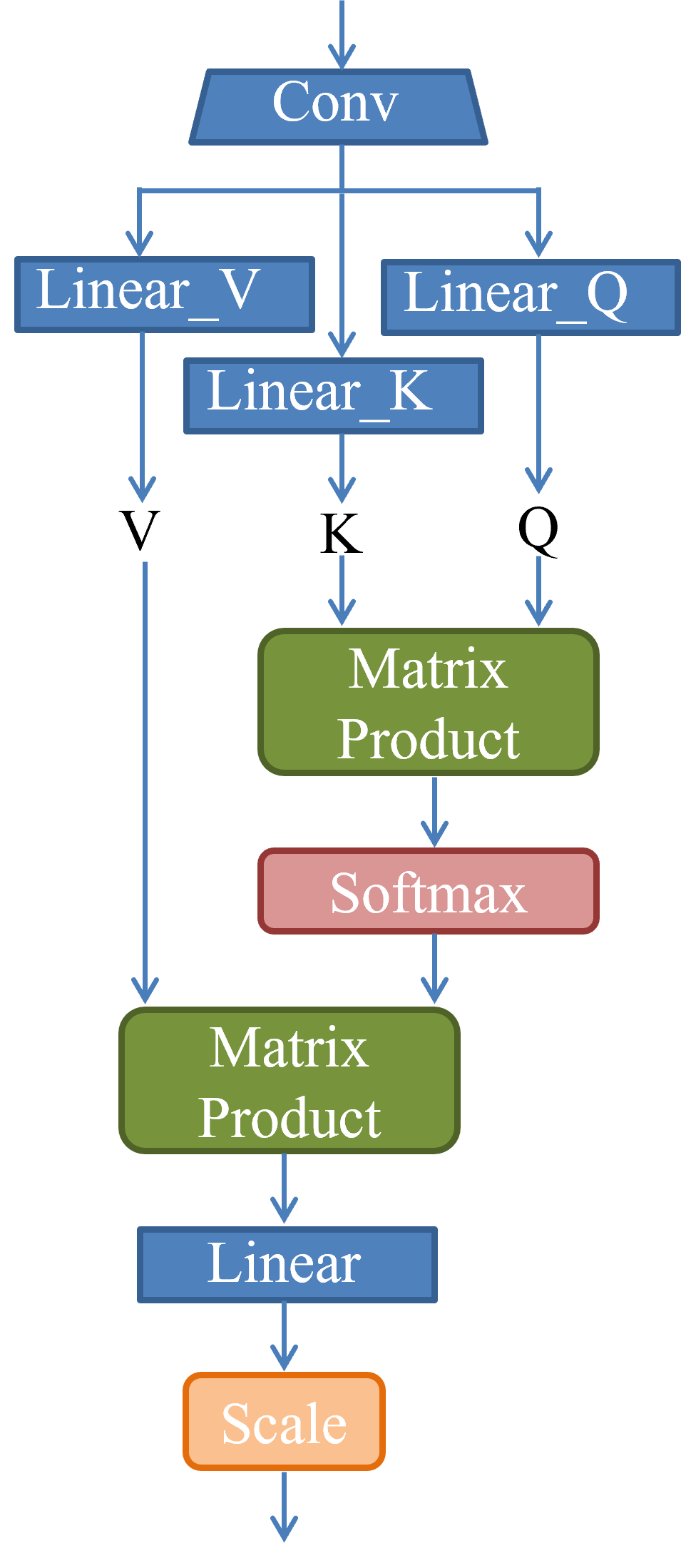}
}
\hspace{40pt}
\subfloat[]{
\includegraphics[scale=0.45,valign=t]{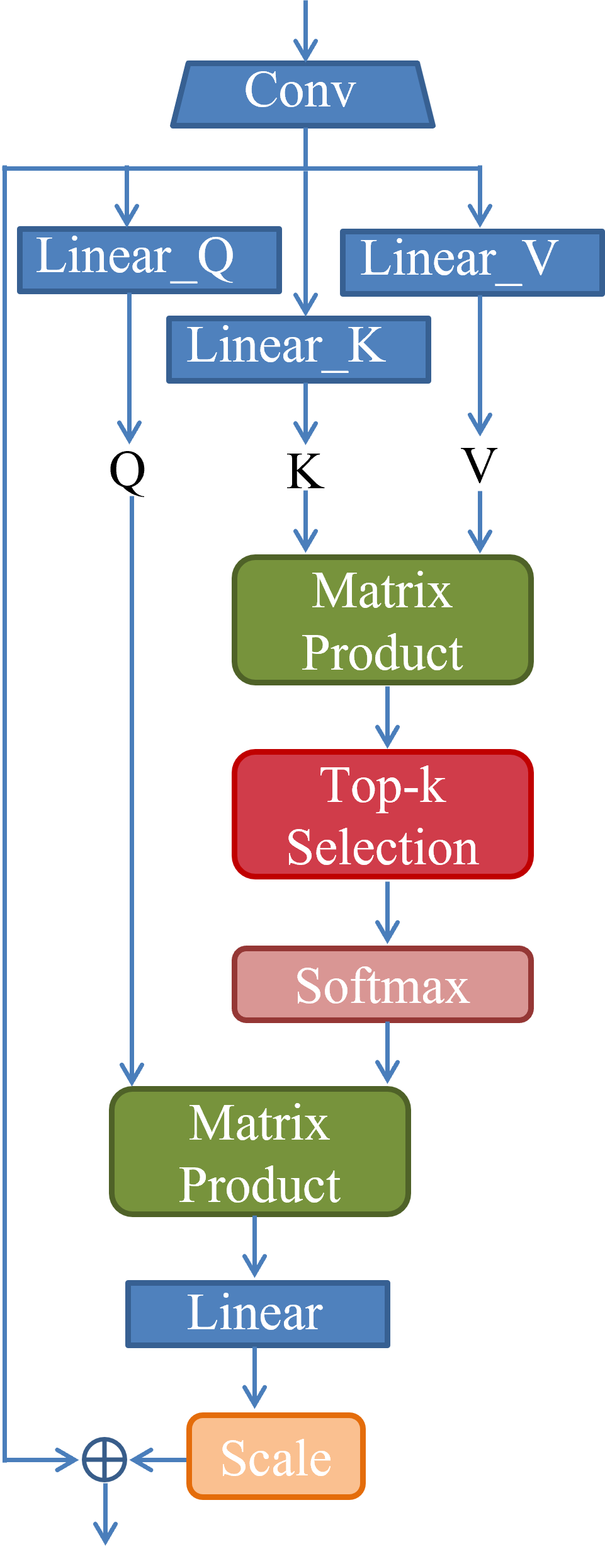}
}
\caption{Self-attention (SA) blocks. (a) Block diagram of the~standard SA block. (b) Block diagram of the proposed efficient residual top-$k$ SA block, where top $k$ selection operator is defined by~(\ref{eq:topk}).}
\label{fig:selfattn}
\end{figure}

 It is desirable to exploit the global receptive field of the~SA module~\cite{vaswani2017attention} for global feature processing. 
However, the vanilla implementation of SA, shown in Fig.~\ref{fig:selfattn}(a), employs all tokens, which injects noise into the training process in high resolution image/video processing tasks by including irrelevant tokens in the computations, thus limiting the performance and increasing the computation burden.
We mitigate this problem by implementing an efficient residual top-$k$ SA (EkSA) implementation, shown in Fig.~\ref{fig:selfattn}(b), which is a modified version of the approach proposed by~\cite{wang2021kvt}.


In the standard SA implementation, each picture sequence with $n$ pixels is mapped to 64 feature channels via regular convolution layers (Conv in Fig.~\ref{fig:selfattn}(a)) for feature extraction. Then, queries (Q $\in \mathbb{R}^{n \times 64} $),  keys (K $\in \mathbb{R}^{n \times 64} $) and values (V~$\in \mathbb{R}^{n \times 64} $) are separately computed by linear mapping layers Linear\_Q, Linear\_K and Linear\_V, where the numbers of output channels of Linear\_Q, Linear\_K and Linear\_V (shown by blue rectangles in Fig.~\ref{fig:selfattn}(a)) are the same as those of the input channels respectively. After obtaining Q, K and V matrices, we multiply Q and K transpose to form the attention weight map, and normalize the result by the softmax operation $\rho(Q \times K^{T})$. Then, we multiply the normalized softmax output and V to obtain an initial attention result. Finally, the initial attention result is mapped through a linear layer and scaled to yield the final attention output. Hence, the standard SA operation can be summarized by:
\begin{flalign}
SA(Q,K,V) = L[\rho(Q \times K^{T}) \times V] \cdot Scale &&
\label{eq:ssa}
\end{flalign}
where T denotes matrix transpose, $\times$ is matrix product, L~represents linear mapping and $Scale$ is a learnable scale factor. The~softmax normalization
\begin{flalign}
\rho(Y) = \sigma_{row}(Y)&&
\label{eq:softmax}
\end{flalign}
is applied row-wise.

Multiple inputs bring in replicated and redundant information since many elements are similar to each other across frames. In order to eliminate irrelevant redundant tokens and accelerate training, we instead implement the top-$k$ SA model~\cite{wang2021kvt} given by:
\begin{flalign}
kSA(Q,K,V) = L[\rho(T_k[Q \times K^{T}]) \times V] \cdot Scale &&
\label{eq:ksa}
\end{flalign}
where  
\begin{flalign}
[T_k(A)]_{ij} = \left\{
\begin{aligned}
A_{ij}, &~&~& if~~A_{ij} \in top-k (row~j), & \\
-\infty, &~&~& \text{otherwise}.&
\end{aligned}
\right.&&
\label{eq:topk}
\end{flalign}
is utilized to select the top $k$ elements (tokens) row-wise. The~selection of $k$ relevant tokens, shown by the red box in Fig.~\ref{fig:selfattn}(b), follows computation of the attention weight map.

While removing irrelevant tokens improves the performance of the model successfully, this vanilla implementation of top-$k$ self-attention is not computationally efficient due to the fact that we only set the irrelevant tokens to zeros instead of removing them thoroughly as~\cite{bolya2022token} and~\cite{li2023svitt}, which decrease the~number of tokens by merging tokens or reducing the length of sequences. In order to improve the computational efficiency significantly, inspired by~\cite{shen2021efficient} and~\cite{xia2022efficient}, we exchange multiplication order by first multiplying V$^T$ and K to generate the attention weight map. Thus, the attention weight map matrix is reduced to 64 $\times$ 64 from n $\times$ n. Hence, the efficient top-$k$ SA (EkSA) can be expressed by:
\begin{flalign}
EkSA(Q,K,V) = L[Q \times \rho(T_k[V^{T} \times K])] \cdot Scale &&
\label{eq:eksa}
\end{flalign}
The efficiency and effectiveness of this top-$k$ SA design, shown in Fig.~\ref{fig:selfattn}(b), are demonstrated in Table~\ref{time} and Table~\ref{kablation}, respectively.

Different from~\cite{wang2021kvt}, we found that adding a residual connection 
between initial features and the scaled output, shown by the plus sign in Fig.~\ref{fig:selfattn}(b), significantly improves the overall test performance. We demonstrate the effectiveness of the residual connection by an ablation in the middle row of Table~\ref{ablation}.

\subsection{Separate reconstruction blocks}
\label{seprecon} 
We observe that using separate reconstruction blocks, called the SepRecon architecture, shown in Fig.~\ref{fig}(a), brings benefits in deinterlacing and demosaicing tasks, where the subsampled inputs have fixed even or odd patterns for the deinterlacing task, and the Bayer subsampling pattern is employed for the demosaicing task. These multiple fixed subsampling patterns enable our SepRecon architecture to learn different model parameters tailored to corresponding field/channel patterns, although the~architecture of each reconstruction module is the same. Hence, high level features can be separately reconstructed according to the specific pattern with the help of the corresponding reconstruction module. In the SepRecon block, each reconstruction module, denoted by Recon, consists of sequential stacking of seven standard Res\_Blocks, expressed by 
\begin{flalign}
&Recon = Res\_Block(Res\_Block(…Res\_Block(integ\_fea)))
\label{eq:seprecon_recon}&
\end{flalign}
where each Res\_Block is depicted in Fig.~\ref{fig}(c), integ\_fea is the~integrated feature tensor (addition of tensors computed by DfConv block and EkSA block).

Specifically, we employ two reconstruction modules in parallel with 7 $\times$ standard Res\_Blocks in each branch (shown by two rows of orange blocks on the right in Fig.~\ref{fig}(a)) for the deinterlacing task to separately reconstruct the missing even and odd fields as formulated by:
\begin{flalign}
&\!\begin{aligned}
&Even\_fea = Recon\_E \\
&Odd\_fea = Recon\_O
\label{eq:seprecon_deinterfea}
\end{aligned}&
\end{flalign}
where Even\_fea and Odd\_fea represent separately reconstructed features for even field and odd field, respectively, Recon\_E and Recon\_O have the same architecture as shown in (\ref{eq:seprecon_recon}) but will learn different parameters for even field and odd field.

Likewise, we employ three reconstruction modules in parallel with 7 $\times$ standard Res\_Blocks in each branch for the demosaicing task to reconstruct each of R, G, and B channels separately expressed by
\begin{flalign}
&\!\begin{aligned}
&R\_fea = Recon\_R &\\
&G\_fea = Recon\_G &\\
&B\_fea = Recon\_B
\label{eq:seprecon_demosfea}
\end{aligned}&
\end{flalign}
where R\_fea, G\_fea and B\_fea represent the separately reconstructed features for R channel, G channel and B channel, respectively, Recon\_R, Recon\_G and Recon\_B have the same architecture as in (\ref{eq:seprecon_recon}) but will learn different parameters for R, G, and B channels.

The SepRecon architecture restores the missing lines and pixel values directionally according to the specific degradation model; thus, producing superior performance compared to a single reconstruction block for both even and odd fields or three color channels. The advantage of the SepRecon architecture is demonstrated by an ablation study in Section~\ref{expr}.

\section{Experimental evaluation} 
\label{expr}
We start by discussing the experimental procedures, such as datasets, evaluation methods, and training details in Section~\ref{setting}. Then, we present comparison of our proposed DfConv+EkSA models with the state-of-the-art both quantitatively in terms of PSNR and SSIM metrics and qualitatively by visual quality in Section~\ref{comp}. Next, we present ablation studies in Section~\ref{abla} to demonstrate the improvements obtained by using each of the proposed modifications such as DfConv blocks, the efficient top-$k$ SA module, etc.. Finally, we show that the generalization performance of our models is superior to competing methods  when trained and tested on different datasets via cross-validation in Section~\ref{gen}.

\subsection{Experimental procedures}
\label{setting}
\noindent\textbf{Datasets.} We selected widely and publicly used datasets which cover different resolutions and amounts of motions to demonstrate our proposed architecture and compare with other methods. We used combinations of UCF101~\cite{soomro2012ucf101}, REDS~\cite{nah2019ntire}, Vimeo-90K~\cite{xue2019video} and Vid4~\cite{sajjadi2018frame} 
sets for training and~testing. 

The UCF101 video dataset (240 $\times$ 320) was separated into training and test sets according to~\cite{soomro2012ucf101} and we tested only on ApplyEyeMakeup test class. 

In REDS video dataset (720 $\times$ 1280), clip00, clip11, clip15 and clip20 (called REDS4) were used for testing and the~remaining clips were used for training \cite{wang2019edvr}.

In Vimeo-90K video dataset (256 $\times$ 448), we utilized the~training and test septuplets as stated in \cite{xue2019video}. 

There is no training set for the Vid4~\cite{sajjadi2018frame} dataset (576 $\times$ 704, 480 $\times$ 704), and we used Vid4 only for testing. We tested all compared methods with their best publicly available trained models for both deinterlacing and demosaicing tasks on Vid4.

In addition, two well-known video clips, \textit{SpongeBob} (480 $\times$ 720) and \textit{Mr. Bean's Holiday} (1920 $\times$ 1080), were used as real-world data to validate the effectiveness of our models.

Interlaced videos are generated by splitting ground-truth progressive videos into odd and even fields and retaining only odd or even fields from alternating frames.

Mosaic video sequences are obtained by keeping pixels in R, G, and B channels according to the Bayer pattern and setting other pixel values to zeros.

Both these acquired interlaced and mosaic sequences are directly input into corresponding proposed models and the immediate scaled outputs are straightforwardly compared without any other preprocessing~\cite{brar2024ai,brar2024detection} or postprocessing steps.

\noindent\textbf{Evaluation Methods.} We use peak signal-to-noise ratio~(PSNR) and structural similarity index measure (SSIM)~\cite{wang2004image} for quantitative evaluation of results. We also provide visuals zooming in on details for subjective evaluation of the results. In addition, temporal profiles~\cite{caballero2017real} is employed to show the temporal consistency by stacking a line of pixels across multiple consecutive frames at fixed width (vertical temporal profiles) or fixed height (horizontal temporal profiles).

\begin{table}[b!]
\caption{Number of parameters for deinterlacing (top) and demosaicing (bottom) models} \vspace{-10pt}
\begin{center}
\begin{tabularx}{\columnwidth}{XXX}
\hline
DfRes & DfRes+SA  
&\textbf{DfConv+EkSA}  \\ \hline \hline
 2,756,739
 &  2,784,276
 &  2,943,235 \\
 \hline
 -
 &  -
 &  3,460,227
\\ \hline
 \end{tabularx} 
\label{parameter}
\end{center}
\end{table}

\begin{table}[b!]
\caption{Test time (sec.) for one output frame with different models, where +EKSA, +kSA and +SA represent DfConv+EkSA, DfConv+kSA and DfRes+SA, respectively (Top: Deinterlacing on UCF101, Bottom: Demosaicing on vid4).}  \vspace{-10pt}
\begin{center}
\begin{tabularx}{\columnwidth}{XXXXXX}
\hline
\textbf{+EkSA} & +kSA & +SA & DfRes & DeT & Ref~\cite{zhu2017real}
\\ \hline
  0.08
 & 0.32
 & 0.17
 & 0.08
 & 0.14
 & 0.09
 \end{tabularx} 
\begin{tabularx}{\columnwidth}{XXXXXX}
\hline \hline
 \textbf{+EkSA}
 & DPIR
 & Kokkinos
 & NTSDCN
 & Ehret
 & BJDD
\\ \hline
  0.65
 & 0.50
 & 5.22
 & 0.97
 & 0.14
 & 0.01
 \\ \hline
 \end{tabularx} 
\label{time}
\end{center}
\end{table}


\noindent\textbf{Training Details.} We input five adjacent fields, $(N-2)_O$, $(N-1)_E$, $N_O$, $(N+1)_E$ and $(N+2)_O$ to estimate field $N_E$, or $(N-2)_E$, $(N-1)_O$, $N_E$, $(N+1)_O$ and $(N+2)_E$ to estimate field $N_O$ for the deinterlacing task, and mosaiced R,~G, B arrays for frames $(N-2)$, $(N-1)$, $N$, $(N+1)$ and $(N+2)$ to restore frame $N$ for the demosaicing task.


We use the loss function:
\begin{flalign}
L = &MSE(Y^{pred},Y^{GT})+0.1 \cdot Char(Y^{pred},Y^{GT})\\
&\nonumber +\lambda_{TV} \cdot TV(Y^{pred})&&
\label{eq:loss}
\end{flalign}
where $Y^{pred}$ and $Y^{GT}$ are estimated and ground truth frames, respectively, $MSE(\cdot)$ is the mean square error, $Char(\cdot)$~denotes the Charbonnier loss~\cite{wang2019edvr}, $TV(\cdot)$ denotes the~total variation regularizer~\cite{zhu2017real}, and $\lambda_{TV}$ is a scale factor that is set to $2.0 \times 10^{-3}$ in all experiments.

We use the Adam optimizer with the initial learning rate $4\times10^{-4}$ and Cosine Annealing scheme, the same as in~\cite{ji2022multi}. We train all models for 150,000 iterations with mini-batch size equal to~32 for deinterlacing and 24 for demosaicing tasks, and the~input patch size is set to $64 \times 80$. 

\begin{figure*}[t]
\centering
\subfloat[\textbf{Ours}]{
   \includegraphics[scale=0.8]{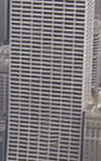}
}
\hspace{-5pt}
\subfloat[DfRes+SA~\cite{ji2022multi}]{
   \includegraphics[scale=0.8]{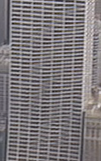}
    \hspace{-5pt}
}
\subfloat[DfRes~\cite{ji2021learned}]{
   \includegraphics[scale=0.8]{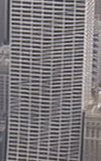}
    \hspace{-5pt}
}
\subfloat[DeT~\cite{song2023transformer}]{
   \includegraphics[scale=0.8]{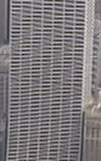}
    \hspace{-5pt}
}
\subfloat[Ref~\cite{zhu2017real}]{
   \includegraphics[scale=0.8]{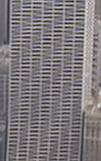}
    \hspace{-5pt}
}
\subfloat[GT]{
   \includegraphics[scale=0.8]{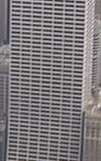}
}
\\

\subfloat[\textbf{Ours}]{
    \includegraphics[scale=2.8]{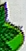}
    \hspace{4.7pt}
}
\subfloat[DPIR~\cite{zhang2021plug}]{
    \includegraphics[scale=2.8]{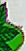}
    \hspace{4.7pt}
}
\subfloat[BJDD~\cite{a2021beyond}]{
    \includegraphics[scale=2.8]{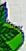}
    \hspace{4.7pt}
}
\subfloat[Kokkinos~\cite{kokkinos2019iterative}]{
    \includegraphics[scale=2.8]{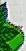}
    \hspace{4.7pt}
}
\subfloat[NTSDCN~\cite{wang2020ntsdcn}]{
    \includegraphics[scale=2.8]{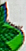}
    \hspace{4.7pt}
}
\subfloat[GT]{
    \includegraphics[scale=2.8]{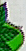}
}
\\

\subfloat[\textbf{Ours}]{
    
    \includegraphics[scale=2.8]{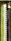}
    \hspace{37pt}
             }
\subfloat[DPIR
]{
    \includegraphics[scale=2.8]{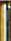}
    \hspace{37pt}
             }
\subfloat[BJDD
]{
    \includegraphics[scale=2.8]{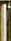}
    \hspace{37pt}
             }
\subfloat[Kokkinos
]{
    \includegraphics[scale=2.8]{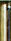}
    \hspace{37pt}
             }
\subfloat[NTSDCN
]{
    \includegraphics[scale=2.8]{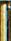}
    \hspace{37pt}
             }
\subfloat[GT]{
    \includegraphics[scale=2.8]{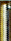}
    \hspace{17pt}
             }
\subfloat[]{
    \includegraphics[scale=0.423]{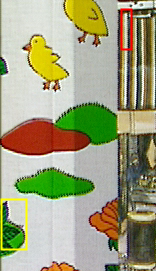}
             }
\caption{(a)-(e) are deinterlaced results on a frame (Vid4) with different models; (f) is the ground truth frame. (g)-(k) and (m)-(q) are crops from a demosaiced frame with different models; (l) and (r) are ground truth crops. The crops (g)-(l) and (m)-(r) correspond to the yellow and red boxes marked on (s).}  
\label{resultsonVid} 
\end{figure*}

\begin{table}[t!]
 \caption{Comparison of models for Video deinterlacing on multiple datasets, where +SA represents DfRes+SA (Top: PSNR, Bottom: SSIM). 
 \vspace{-6pt}
 }
\begin{center}
\begin{tabularx}{\columnwidth}{X|XXXXX}
\hline
 & \textbf{Ours} 
 & +SA 
 & DfRes 
 & DeT~\cite{song2023transformer}
 & Ref\cite{zhu2017real} 
 \\\hline \hline
\multirow{2}*{UCF101}
 & 52.34
 & 51.92
 & 51.48
 & 43.46
 & 36.96
 \\ 
 & 0.9994
 & 0.9993
 & 0.9993
 & 0.9951
 & 0.9859
 \\  \hline
 
 \multirow{2}*{Vimeo}
 & 44.96
 & 44.35
 & 43.45
 & 42.22
 & 38.81
 \\ 
 & 0.9956
 & 0.9949
 & 0.9935
 & 0.9850
 & 0.9740
 \\  \hline
 
 \multirow{2}*{REDS}
 & 38.04
 & 37.00
 & 36.51
 & 34.08
 & 33.45
 \\
 & 0.9870
 & 0.9835
 & 0.9817
 & 0.9513
 & 0.9455
 \\  \hline
\end{tabularx}
\label{on3dataset}
\end{center}  \vspace{-5pt}
\end{table}

\begin{table}[h!]
\caption{Comparison of models for Video demosaicing on Vid4 (above double lines) and on REDS4 (below double lines), respectively (Top: PSNR, Bottom: SSIM).}  \vspace{-6pt}
\begin{center}
\begin{tabularx}{\columnwidth}{XXXXX}
\hline
 \textbf{Ours}
 & DPIR 
 & BJDD
 & Kokkinos 
 & NTSDCN 
 \\ \hline \hline
 39.44
 & 37.91
 & 37.55
 & 34.05
 & 32.77
 \\
 0.9897
 & 0.9816
 & 0.9819
 & 0.9834
 & 0.9868
  \\  \hline \hline
  48.04
 & 43.89
 & 44.54
 & 32.37
 & 36.51
\\ 
 0.9982
 & 0.9895
 & 0.9918
 & 0.9534
 & 0.9975
 \\  \hline
\end{tabularx}
\label{demosaiconvid}
\end{center} 
\end{table}

\subsection{Comparison with the state-of-the-art}
\label{comp} 
We compare our proposed DfConv+EkSA model with DfRes~\cite{ji2021learned}, DfRes+SA~\cite{ji2022multi}, DeT~\cite{song2023transformer}, Gao~{\it et al.}~\cite{gao2023revitalizing} and Zhu {\it et al.}~\cite{zhu2017real} for the~deinterlacing task, and with DPIR~\cite{zhang2021plug}, BJDD~\cite{a2021beyond}, Kokkinos \cite{kokkinos2019iterative}, NTSDCN \cite{wang2020ntsdcn} and Ehret~\cite{ehret2019joint} for the~demosaicing task, where Zhu {\it et al.}~\cite{zhu2017real}, DPIR, BJDD and NTSDCN are designed for single picture input, and comparison with Gao~{\it et al.}~\cite{gao2023revitalizing} is only literally stated in Section~\ref{gen} due to unavailability of codes. We also provide comparisons with the classical video superresolution architectures (EDVR~\cite{wang2019edvr}, TDAN~\cite{tian2020tdan}, DUF~\cite{jo2018deep}) trained on our deinterlacing video datasets. In line with the availability of training and test codes of different compared methods, we retrain DfRes, DfRes+SA, EDVR, TDAN and DUF with the same training datasets, and adopt the best available pre-trained models of other methods for validation. 

All the models implement a deterministic mapping from input (interlaced or mosaic) images to output (deinterlaced or demosaiced) images, not to be confused with stochastic generative models that provide a different result depending on the seed value. Potential differences between repeated experiments are limited to any precision difference between different computing platforms and can be ignored. The statistical significance of the results may be evaluated by running the trained models across multiple benchmark training/ test datasets, which we did in the paper. Our extensive experimental results with different combinations of training and test sets clearly show the superior performance of our models.


As a measure of model complexity, the numbers of parameters in the compared models are very close to each other as shown in Table~\ref{parameter}. Table~\ref{time} shows the computational speed (test run time) competitiveness of our DfConv+EkSA model.

\begin{figure}[t!]
\centering
\subfloat[\textbf{Ours}]{
   \includegraphics[scale=1.7]{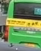}
    \hspace{-5pt}
}
\subfloat[BJDD~\cite{a2021beyond}]{
   \includegraphics[scale=1.7]{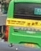}
    \hspace{-5pt}
}
\subfloat[DPIR~\cite{zhang2021plug}]{
   \includegraphics[scale=1.7]{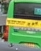}
}
\\
\subfloat[NTSDCN~\cite{wang2020ntsdcn}]{
   \includegraphics[scale=1.7]{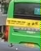}
    \hspace{-5pt}
}
\subfloat[Kokkinos~\cite{kokkinos2019iterative}]{
   \includegraphics[scale=1.7]{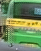}
    \hspace{-5pt}
}
\subfloat[Mosaic]{
   \includegraphics[scale=1.7]{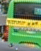}
}
\\
\subfloat[\textbf{Ours}]{
   \includegraphics[scale=1.7]{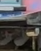}
    \hspace{-5pt}
}
\subfloat[BJDD~\cite{a2021beyond}]{
   \includegraphics[scale=1.7]{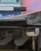}
    \hspace{-5pt}
}
\subfloat[DPIR~\cite{zhang2021plug}]{
   \includegraphics[scale=1.7]{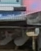}
}
\\
\subfloat[NTSDCN~\cite{wang2020ntsdcn}]{
   \includegraphics[scale=1.7]{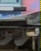}
    \hspace{-5pt}
}
\subfloat[Kokkinos~\cite{kokkinos2019iterative}]{
   \includegraphics[scale=1.7]{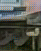}
    \hspace{-5pt}
}
\subfloat[Mosaic]{
   \includegraphics[scale=1.7]{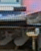}
}
\\
\subfloat[]{
   \includegraphics[scale=0.8]{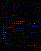}
    \hspace{-4pt}
}
\subfloat[]{
   \includegraphics[scale=0.8]{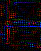}
    \hspace{-4pt}
}
\subfloat[]{
   \includegraphics[scale=0.8]{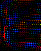}
    \hspace{-4pt}
}
\subfloat[]{
   \includegraphics[scale=0.8]{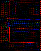}
    \hspace{-4pt}
}
\subfloat[]{
   \includegraphics[scale=0.8]{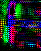}
    \hspace{-4pt}
}
\subfloat[]{
   \includegraphics[scale=0.8]{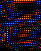}
}
\\
\subfloat[]{
   \includegraphics[scale=0.8]{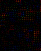}
    \hspace{-4pt}
}
\subfloat[]{
   \includegraphics[scale=0.8]{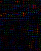}
    \hspace{-4pt}
}
\subfloat[]{
   \includegraphics[scale=0.8]{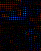}
    \hspace{-4pt}
}
\subfloat[]{
   \includegraphics[scale=0.8]{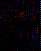}
    \hspace{-4pt}
}
\subfloat[]{
   \includegraphics[scale=0.8]{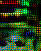}
    \hspace{-4pt}
}
\subfloat[]{
   \includegraphics[scale=0.8]{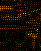}
}
\caption{More demosaicing results on REDS4 data. (f) and (l) are mosaic frames. (a)-(e) and (g)-(k) are demosaiced results. (m)-(x) are 10 times pixel differences between (a)-(l) and GT, respectively.
}
\label{DMreds4} 
\end{figure}

\vspace{4pt}
\noindent\textbf{Quantitative Performance Comparison.} 
Quantitative comparison results for the deinterlacing and demosaicing tasks are provided in Table \ref{on3dataset} and Table \ref{demosaiconvid}, respectively.

Table~\ref{on3dataset} shows that our proposed DfConv+EkSA model with the~new offset estimation and efficient residual top-$k$ self attention modules trained for the deinterlacing task is clearly superior to competing methods on multiple datasets. Specifically, Table~\ref{on3dataset} shows +2.7 dB gain on Vimeo, +4 dB gain on REDS and +9 dB gain on UCF101 compared to DeT~\cite{song2023transformer}. Although narrower gains are reported compared to DfRes and DfRes+SA, which are earlier versions of our model presented in short conference papers~\cite{ji2022multi,ji2021learned}. Yet, Table~\ref{on3dataset} still shows a gain of 0.42 dB on UCF101, 0.61 dB on Vimeo and 1.04 dB on REDS compared to DfRes+SA~\cite{ji2022multi}, which is a significant gain compared to our earlier work. Hence, our results are the state-of-the-art in the~deinterlacing task.

Table~\ref{demosaiconvid} shows that our DfConv+EkSA model trained for the demosaicing task also outperforms the state-of-the-art demosaicing methods on both Vid4 and REDS4 datasets. Hence, it is the state-of-the-art in the~demosaicing task by a large margin.
\vspace{4pt}

\begin{figure}[t!]
\centering
\subfloat[\textbf{Ours}]{
   \includegraphics[scale=1]{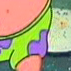}
    \hspace{-5pt}
}
\subfloat[DeT~\cite{song2023transformer}]{
   \includegraphics[scale=1]{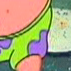}
    \hspace{-5pt}
}
\subfloat[DfRes+SA~\cite{ji2022multi}]{
   \includegraphics[scale=1]{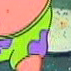}
}
\\
\subfloat[DfRes~\cite{ji2021learned}]{
   \includegraphics[scale=1]{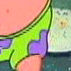}
    \hspace{-5pt}
}
\subfloat[Ref~\cite{zhu2017real}]{
   \includegraphics[scale=1]{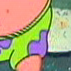}
    \hspace{-5pt}
}
\subfloat[Interlaced]{
   \includegraphics[scale=1]{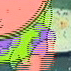}
}
\\
\subfloat[\textbf{Ours}]{
    \includegraphics[scale=0.363]{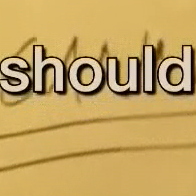}
    \hspace{-5pt}
}
\subfloat[DeT~\cite{song2023transformer}]{
    \includegraphics[scale=0.363]{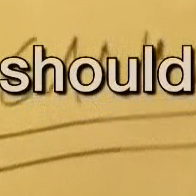}
    \hspace{-5pt}
}
\subfloat[DfRes+SA~\cite{ji2022multi}]{
    \includegraphics[scale=0.363]{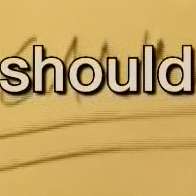}
}
\\
\subfloat[DfRes~\cite{ji2021learned}]{
    \includegraphics[scale=0.363]{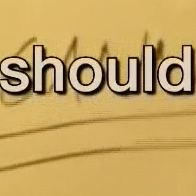}
    \hspace{-5pt}
}
\subfloat[Ref~\cite{zhu2017real}]{
    \includegraphics[scale=0.363]{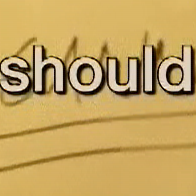}
    \hspace{-5pt}
}
\subfloat[Interlaced]{
    \includegraphics[scale=0.363]{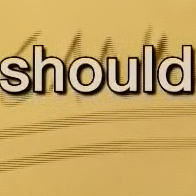}
}

\caption{Deinterlacing results on real-world data. (f) and (l) are interlaced frames, where even and odd fields are weaved. (a)-(e) and (g)-(k) are deinterlaced results for the bottom field of the corresponding interlaced frames.
}  
\label{realworld} 
\end{figure}

\begin{figure*}[h!]
\begin{center}
\subfloat[Horizontal temporal profiles on calendar clip (74 frames).]{
   \includegraphics[scale=0.8]{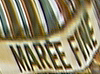}\hspace{0pt}
   \includegraphics[scale=0.8]{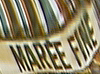}\hspace{0pt}
   \includegraphics[scale=0.8]{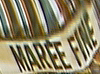}\hspace{0pt}
   \includegraphics[scale=0.8]{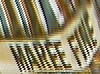}\hspace{0pt}
   \includegraphics[scale=0.8]{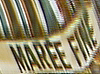}\hspace{0pt}
}
\\
\subfloat[Vertical temporal profiles on calendar clip (74 frames).]{
   \includegraphics[scale=1]{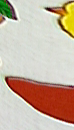}\hspace{7.5pt}
   \includegraphics[scale=1]{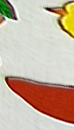}\hspace{7.5pt}
   \includegraphics[scale=1]{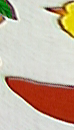}\hspace{7.5pt}
   \includegraphics[scale=1]{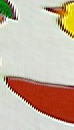}\hspace{7.5pt}
   \includegraphics[scale=1]{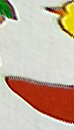}\hspace{0.5pt}
}
\\
\subfloat[Horizontal temporal profiles on city clip (60 frames).]{
   \includegraphics[scale=0.8]{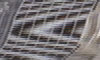}\hspace{0pt}
   \includegraphics[scale=0.8]{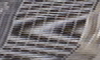}\hspace{0pt}
   \includegraphics[scale=0.8]{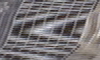}\hspace{0pt}
   \includegraphics[scale=0.8]{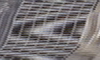}\hspace{0pt}
   \includegraphics[scale=0.8]{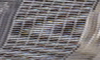}\hspace{0pt}
}
\\
\subfloat[Vertical temporal profiles on city clip (60 frames).]{
   \includegraphics[scale=1]{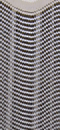}\hspace{25pt}
   \includegraphics[scale=1]{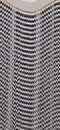}\hspace{25pt}
   \includegraphics[scale=1]{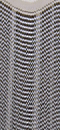}\hspace{25pt}
   \includegraphics[scale=1]{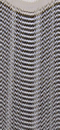}\hspace{25pt}
   \includegraphics[scale=1]{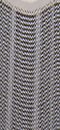}\hspace{0pt}
}
\caption{Temporal profiles on full-frame-rate calendar and city clips of Vid4 dataset for deinterlacing task, from left to right: DfConv\_EkSA, DfRes\_SA, DfRes, DeT and Ref~\cite{zhu2017real}.
}  
\label{temporalprofiles} 
\end{center}
\end{figure*}

\noindent\textbf{Visual Performance Comparison.} 
It is generally accepted that PSNR/SSIM results do not always correlate well with visual evaluations. To this effect, we also provide visual comparisons. Fig.~\ref{resultsonVid}~(a)-(e) show deinterlaced frames with different models. Our method produces almost no stripe distortions compared to other four methods.

Fig.~\ref{resultsonVid}~(g)-(k) and (m)-(q) show demosaiced frames with different models. Close inspection of (g)-(k) show that our DfConv+EkSA model produces less color distortion and (m)-(q) show that our method generates sharper images compared to other methods which look over-smoothed due to denoising.

Fig.~\ref{DMreds4} shows more visual results for the demosaicing task on REDS4 dataset. We can see that (c) and (i) bring in slight vertical strip noise when zoomed in on plate and white area; (d) and (j) produce color distortion while dealing with edge information; while (e) and (k) fail to demosaic frames successfully. Due to the pixel level operation in demosaicing, it is hard to visually tell the differences between (a) and (b), or (g) and (h). However, with the help of difference images from (m) to (x), we can easily find that our method produces less differences from ground truth frames as shown in (m) and (s). In overall contrast, our proposed method demosaics input frames well avoiding generating strip noise or color distortion as shown in (a), (g), (m) and (s).

Fig.~\ref{realworld} shows visual results on two real-world video clips. Our proposed method (a) can relieve severe comb-teeth caused by interlacing (f) compared with other methods through (b)-(e). In addition, we can see that (i) and (j) effectively preserve the letters but generate artificial patterns with the horizontal lines; (k) deinterlaces the lines well but yields white dot noise on the top arcs of letters s, o and d, and fades top lines of letters h, l and d. Instead, our proposed model can deinterlace both the letters and the horizontal lines properly without producing additional noise. DeT~\cite{song2023transformer} deinterlaces (h) well with indistinguishable visual effect compared to ours (g). This is because of the simple texture pattern of (h) in contrast to more complex texture patterns in (b) and Fig.~\ref{resultsonVid}~(d) where DeT introduces noises.

In summary, visual results demonstrate that our proposed model keeps genuine details well without introducing extra noise or artificial patterns in a variety of video clips.
\vspace{4pt}

\noindent\textbf{Temporal Consistency Comparison.} 
Temporal consistency of reconstructed frames is usually evaluated by a temporal profile, which refers to a plot of a fixed line from each frame throughout the duration of video. Discontinuity or jaggedness in the~vertical direction in the temporal profile is an indicator of lack of temporal consistency.
Fig.~\ref{temporalprofiles} shows the temporal profiles of different deinterlacing methods. Deformable convolution based methods (left three images), in terms of both horizontal and vertical profiles, have superior temporal consistency compared to DeT~\cite{song2023transformer} and Ref~\cite{zhu2017real} on both clips. Moreover, on city clip (c) and~(d) with more complex pattern, the proposed DfConv\_EkSA method produces smoother profiles and less stripe noise under close inspection.


 
 

\begin{table*}[t!]
\caption{Ablation studies for the deinterlacing task: Top row: Comparison of DfConv vs. DfRes blocks, and integration of features by addition $+$ vs.  concatenation $|$. Middle row: Comparison of $k$SA vs. SA module for $k=100$ with and without residual connection. Bottom row: Comparison of separate reconstruction modules with 7 Res\_Blocks each vs. single module with 0, 7, 14 Res\_Blocks.}   \vspace{-10pt}
\begin{center}
\begin{tabularx}{\textwidth}{XX|XXXX}
\hline
 \multicolumn{2}{c|}{Feature Alignment and Integration} & DfConv+SA & Df+SA & DfRes+SA & DfRes$|$SA \\
 \hline 
 \multicolumn{2}{c|}{PSNR}
 & 52.22
 & 52.16
 & 51.92
 & 51.80
 \\ 
 \multicolumn{2}{c|}{SSIM}
 & 0.99934
 & 0.99933
 & 0.99931
 & 0.99931
  \\  \hline\hline
 \multicolumn{2}{c|}{Self-Attention Block} & $k$SA $w\backslash$ res & $k$SA $wo\backslash$ res & SA $w\backslash$ res & SA $wo\backslash$ res \\
 \hline 
 \multicolumn{2}{c|}{PSNR}
 & 52.32
 & 51.52
 & 52.22
 & 51.56
 \\ 
 \multicolumn{2}{c|}{SSIM}
 & 0.99936
 & 0.99924
 & 0.99934
 & 0.99922
 \\  \hline\hline
 \multicolumn{2}{c|}{Reconstruction Stage}
 & SepRecon
 & 0
 & 7
 & 14
 \\   \hline
 \multicolumn{2}{c|}{PSNR}
 & 51.92 
 & 50.98
 & 51.37
 & 51.23
 \\
 \multicolumn{2}{c|}{SSIM}
 & 0.99931 
 & 0.99921
 & 0.99923
 & 0.99917
 \\   \hline
  
\end{tabularx}
\label{ablation}
\end{center}
\end{table*}

\begin{table}[t!]
\caption{Ablation study on the value of $k$ for top-k self attention (Top: PSNR, Bottom: SSIM).}  \vspace{-10pt}
\begin{center}
\begin{tabularx}{\columnwidth}{XXXX}
\hline
\multicolumn{4}{c}{$k$ values for DfConv+kSA } 
\\
\textbf{50} & 100 &150 &all
\\ \hline
 52.34
 & 52.32
 & 52.30
 & 52.22
 \\
 0.99936
 & 0.99936
 & 0.99935
 & 0.99934
\\ \hline \hline
\multicolumn{4}{c}{$k$ values for DfConv+EkSA} 
\\
32 & \textbf{50} & 64
&-
\\ \hline
 52.32
 & 52.34
 & 52.28
 & -
 \\
 0.99936
 & 0.99936
 & 0.99936
 & -
 \\ \hline
\end{tabularx} 
\label{kablation}
\end{center}
\end{table}

\begin{table*}[pt!]
\caption{Generalization performance of our  DfConv+EkSA model trained and tested on different datasets for deinterlacing and demosaicing tasks (Top: PSNR, Bottom: SSIM).}   \vspace{-10pt}
\begin{center}
\begin{tabularx}{\textwidth}{X|X XXX|XXX}
\hline
  {\multirow{2}{*}{Train on}}
   & 
   & \multicolumn{3}{c|}{\textbf{Deinterlacing:} Test on } 
   & \multicolumn{3}{c}{\textbf{Demosaicing:} Test on } \\ 
   & 
   & UCF101 & Vimeo-90K & REDS4 
   & UCF101 & Vimeo-90K & REDS4
   \\ \hline\hline
\multirow{2}{*}{UCF101} 
 & PSNR
 & 52.34
 & 41.49
 & 35.40
 & 52.64
 & 43.91
 & 44.50
  \\  
 & SSIM
 & 0.9994
 & 0.9917
 & 0.9771
 & 0.9994
 & 0.9944
 & 0.9967
 \\ \hline
\multirow{2}{*}{Vimeo-90K} & PSNR
 & 47.57
 & 44.96
 & 36.67
 & 46.99
 & 48.01
 & 45.76
 \\  
 & SSIM
 & 0.9985
 & 0.9956
 & 0.9824
 & 0.9980
 & 0.9975
 & 0.9974
 \\ \hline
\multirow{2}{*}{REDS} & PSNR
 & 38.71
 & 43.00
 & 38.04
 & 44.64
 & 43.69
 & 48.04
 \\  
 & SSIM
 & 0.9932
 & 0.9941
 & 0.9870
 & 0.9968
 & 0.9942
 & 0.9982
\\ \hline
\multirow{2}{*}{YOUKU} & PSNR
 & 48.73
 & 40.14
 & 34.38
 & 50.18
 & 44.06
 & 44.61
 \\  
 & SSIM
 & 0.9989
 & 0.9890
 & 0.9725
 & 0.9991
 & 0.9948
 & 0.9968
\\ \hline
 \end{tabularx}
\label{generalization}
\end{center}  
\end{table*}

\begin{table*}[t!]
\caption{Comparative generalization results for the deinterlacing task tested on Vid4 dataset (Top: PSNR, Bottom: SSIM). } 
\begin{center}
\begin{tabularx}{\textwidth}{X|XXXXXXX}
\hline
Train on &\textbf{Ours}&DfRes+SA~\cite{ji2022multi}&DfRes~\cite{ji2021learned} &$\Delta$DfRes~\cite{ji2021learned} &EDVR &TDAN
 &DUF~\cite{jo2018deep}
 \\\hline  \hline
\multirow{2}*{UCF101}  
 & 34.61
 & 34.17
 & 33.62
 & 33.23
 & 33.35
 & 32.27
 & 32.43
 \\ 
 & 0.9790
 & 0.9771
 & 0.9751
 & 0.9712
 & 0.9738
 & 0.9687
 & 0.9691
 \\  \hline
 
 \multirow{2}*{Vimeo-90k}
 & 34.85
 & 34.49
 & 33.60
 & 33.93
 & 33.69
 & 33.32
 & 33.46
 \\ 
 & 0.9795
 & 0.979
 & 0.9753
 & 0.9760
 & 0.9756
 & 0.974
 & 0.9748
 \\  \hline
 
 \multirow{2}*{REDS}
 & 34.71
 & 34.12
 & 33.58
 & 33.56
 & 30.39
 & 26.47
 & 31.44
 \\
 & 0.9785
 & 0.9777
 & 0.9756
 & 0.9758
 & 0.9525
 & 0.9148
 & 0.9650
  \\  \hline 
 
\end{tabularx}
\label{generability}
\end{center}  
\end{table*}

\subsection{Ablation studies}
\label{abla}
We conducted the following ablation studies to demonstrate the~effect of each component of the proposed architecture shown in Fig.~\ref{fig} on the~performance of our model: i)~the~structure of new DfConv blocks,  ii)~the~use of top-$k$ SA vs. SA block both with or without residual connection, iii)~the~effect of different values of $k$ on the performance of top-$k$ SA, iv)~combining features aligned by DfRes and SA through addition vs. concatenation,
v)~the~structure of the~reconstruction module. The~results of these ablations on the UCF101 dataset, summarized in Table~\ref{ablation} and Table~\ref{kablation}, are discussed in detail in the following:

i) The structure of new DfConv blocks, i.e, estimating offsets by using DfConv vs. DfRes~\cite{ji2022multi} and using two Df layers with a convolution layer in between vs. only two Df layers. In~our prior work~\cite{ji2022multi}, we show that DfRes offset estimation, which applies deformable convolution residual block only to the reference picture, yields performance improvement compared to the~offset estimation used in TDAN~\cite{tian2020tdan} and EDVR~\cite{wang2019edvr}, which applies regular convolution layers to concatenated neighboring and reference pictures. In~this paper, we show, in the top part of Table~\ref{ablation}, that about 0.3 dB further PSNR improvement can be obtained by our new DfConv offset estimation method compared to DfRes offset estimation.
Moreover, the top row of Table~\ref{ablation} shows that Df block without a residual connection (Df+SA) yields better performance than DfRes block with a residual connection (DfRes+SA), and that Df layers with a convolution layer in between (the proposed DfConv+SA) produces the best performance.

ii)~The use of $k$SA vs. SA block both with or without residual connection:
Then, in the middle row of Table~\ref{ablation}, we compare top-$k$ Self Attention (kSA, where $k$ is set equal to 100) and vanilla SA modules, and both with residual connection and without residual connection. Results indicate that a residual connection in both $k$SA and SA modules can significantly improve performance and $k$SA is better than SA.

iii) The effect of different values of $k$ on the performance of top-$k$ SA: Table~\ref{kablation} shows $k=50$ is the best value for both DfConv+kSA model which utilizes the normal multiplication order as depicted in Eq.~\ref{eq:ksa}, and DfConv+EkSA model which changes the multiplication order as given in Eq.~\ref{eq:eksa}. Moreover, this table also shows that changing multiplication order does not affect the performance when $k$ is set equal to 50 for top-$k$ SA, which is different from standard SA~\cite{ji2022multi}.

iv) Combining features aligned by DfRes and SA through addition vs. concatenation:
The paper \cite{ji2022multi} also shows that the features computed by the SA module complements the features aligned by the deformable convolution blocks to yield significantly improved performance. While SA alone does not provide good performance, it yields significantly improved performance when combined with the features computed by the deformable convolution blocks. We show, at the right part of the top row in Table~\ref{ablation} that combining deformable convolution blocks and SA module by addition (DfRes+SA) is superior to concatenation (DfRes$|$SA). 

v) The structure of the reconstruction module: We investigate the improvement that comes from using the SepRecon module instead of a single reconstruction module with 7 or 14 residual blocks or no residual blocks at all. The bottom row of Table~\ref{ablation} shows that there is more than 0.5 dB gain in using the SepRecon module with 7 residual blocks in each branch.


\subsection{Evaluation of generalization performance}
\label{gen} 

In order to evaluate the generalization ability as to different datasets of our architecture, we train different models on the training sets UCF101, REDS, Vimeo-90k, and YOUKU \cite{YOUKU}, and cross-test them on other testsets. Table~\ref{generalization} shows that our models generalize well to different testsets but their performance may vary according to the amount of motion present in the training and test sets. 

We also trained super-resolution based methods adapted for deinterlacing according to the comparison strategy in \cite{ji2022multi} on three different datasets and only tested on the Vid4 dataset~\cite{sajjadi2018frame} to evaluate the best training dataset (out of three) for the deinterlacing task. Table~\ref{generability} shows that the~PSNR and SSIM generalization performance of all models (except one) are the best when they are trained on the~Vimeo dataset. This table also demonstrates the high training stability (less fluctuation of PSNR and SSIM) of our model on different training datasets compared to other methods on these three training datasets. Moreover, our method has an overall performance superiority over all other methods.

SSIM gains are not as big as the PSNR gains reported in Tables~\ref{on3dataset}-\ref{generability}. This is mainly because the highest possible value for SSIM is 1 and it is getting more difficult to report larger gains as the results approach that limit. Although the SSIM gains are small, our proposed method DfConv+EkSA still outperforms our prior models DfRes+SA and DfRes on all datasets, which have been ranked 1 and 2 at the time of submission in the MSU Deinterlacer Benchmark~\cite{MSU} in terms of both PSNR and mean opinion score (MOS). Gains over other methods in Tables~\ref{on3dataset}-\ref{generability} are relatively bigger given that the largest SSIM value is 1.

It is worth noting that the SSIM results tested on Vimeo test set trained on Vimeo and REDS datasets shown in Table~\ref{generalization} are higher than those tested on Vimeo dataset in the second table of the recent deinterlacing paper~\cite{gao2023revitalizing}. Moreover, the SSIM results tested on Vid4 dataset of our proposed methods shown in Table~\ref{generability} are also higher than those tested on Vid4 dataset in~\cite{gao2023revitalizing}.

Extensive experimental results reveal that both training and test datasets affect the performance of models for both tasks, i.e., training the same network on different training datasets produces different performances on the same testset, and a model trained on a single training dataset produces different performances on different testsets.


\section{Conclusion}
\label{conc}
\par We introduce a new learned model, called DfConv+EkSA, that achieves the state-of-the-art performance in both video deinterlacing and demosaicing tasks. The~proposed design consists of two main stages: namely, feature alignment and integration stage and separate reconstruction stage, each containing novelties that contribute to the performance improvements as demonstrated by the ablation studies. 
The feature alignment and integration stage combines local spatio-temporal features aligned by novel modified deformable convolution (DfConv) blocks and global features processed by a novel efficient top-$k$ self-attention (EkSA) block additively, in order to benefit from both local and non-local video features. In the~reconstruction stage, we use separate reconstruction modules (7 blocks each) for different type of pictures, which outperforms using a single reconstruction module with 7 or 14 blocks. The model design choices we made are justified by ablation studies.

Extensive experimental results demonstrate that the proposed model with the proposed design novelties yields the~state-of-the-art performance and generalization ability (across different datasets) for both tasks. This work demonstrates that the proposed learned demosaicing and deinterlacing models should be an integral part of consumer video production and processing pipelines. The source code and models will be made publicly available after the~review process.

\vspace{10pt}
\noindent\textbf{Acknowledgments}
\vspace{10pt}

This work is supported by TUBITAK 2247-A Award No. 120C156 and KUIS AI Center funded by Turkish Is Bank. AMT acknowledges Turkish Academy of Sciences. RJ acknowledges support from Fung Foundation.



\bibliographystyle{elsarticle-num} 
\bibliography{IMAVIS}





\end{document}